\documentclass[useAMS,usenatbib]{mn2e}
\usepackage{times}
\usepackage{epsfig}
\usepackage{colordvi}
\usepackage{graphicx}
\usepackage{amssymb}

\def\apj{ApJ}
\def\apjl{ApJ}
\def\apjs{ApJS}
\def\aap{A\&A}
\def\mnras{MNRAS}
\def\rmd{{\rm d}}
\def\source{{3C~454.3}}
\def\sigmaln{{\sigma_{\rm ln}}}

\title[Gamma-ray emission of 3C 454.3]
{Variation of the $\gamma\gamma$ opacity by the \mbox{He\,{\sc ii}} Lyman continuum 
constrains the location of the $\gamma$-ray emission region  in the blazar 3C~454.3}

\author[B. E. Stern and J. Poutanen]
    {Boris E. Stern$^{1,2}$\thanks{E-mail: stern.boris@gmail.com, juri.poutanen@oulu.fi}   and     Juri~Poutanen$^{3}$\footnotemark[1] \\
$^1$Institute for Nuclear Research, Russian Academy of Sciences, Prospekt 60-letiya Oktyabrya 7a, Moscow 117312, Russia\\
$^2$Astro Space Center, Lebedev Physical Institute, Profsoyuznaya 84/32,  Moscow 117997, Russia \\
$^3$Astronomy Division, Department of Physics, PO Box 3000, FIN-90014 University of Oulu, Finland }

\begin{document}
%\date{}
\date{Accepted 2011 June 29. Received 2011 June 29; in original form 2011 May 13}
\pagerange{\pageref{firstpage}--\pageref{lastpage}} \pubyear{2011}

\maketitle

\label{firstpage}

\begin{abstract} 
We study spectral properties of the brightest $\gamma$-ray  blazar 3C~454.3 using 138 weeks of observations 
by the {\em Fermi Gamma-ray Space Telescope (Fermi)}. 
We probe the behaviour of the source as a function of time at different brightness levels. 
The {\em Fermi} spectra in the GeV  range can be well described by a wide underlying lognormal distribution with the 
photon-photon absorption breaks produced by  the \mbox{He\,{\sc ii}} and \mbox{H\,{\sc i}}  Lyman 
recombination continua (LyC). 
We find a power-law dependence of the peak energy on flux 
and discover anti-correlation between  
the column density of the \mbox{He\,{\sc ii}}  LyC and flux. 
This implies that the $\gamma$-ray emission zone lies close to the boundary of the high-ionization part of the 
broad-line region and moves away from the black hole when the flux increases. 
Identification of the $\gamma$-ray production with the relativistic jet, 
implies that the jet is already accelerated at sub-parsec distances from the central black hole,
which favours the Blandford-Znajek process as the jet launching mechanism. 
\end{abstract}

%________________________________________________________________

\section{Introduction}

\begin{figure*}
\centerline{ \epsfig{file=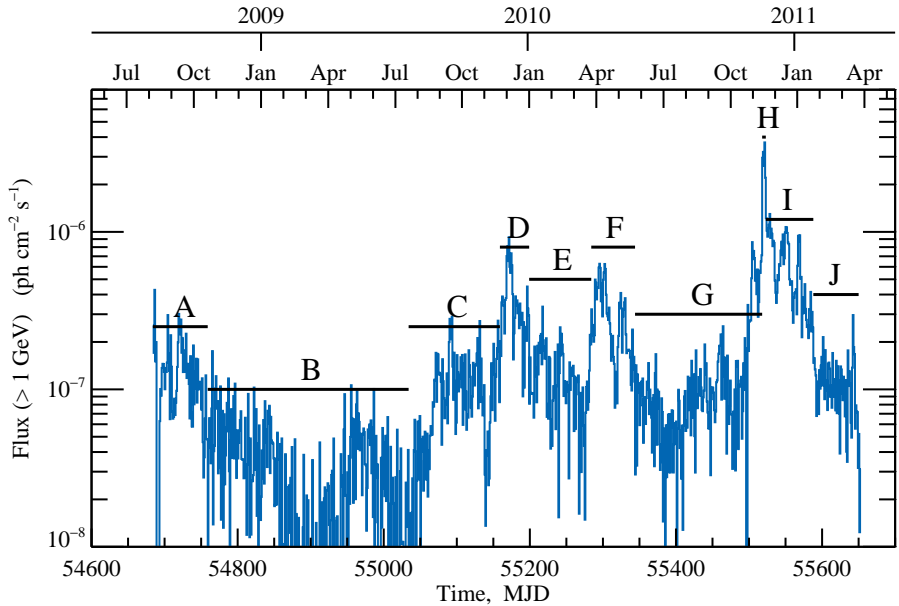, width=12cm}}
\caption{The light curve of the  $\gamma$-ray 
blazar \source\ from the beginning of the {\em Fermi} data records to 2011 April 1. 
Horizontal bars mark the time intervals used in the spectral analysis (see Table \ref{tab:results}).}
\label{fig:lc}
\end{figure*}

The flat spectrum radio quasar (FSRQ) \source\ at redshift $z=0.859$ is by far the brightest blazar in the $\gamma$-ray range during the lifetime  
of the {\em Fermi Gamma-ray Space Telescope} ({\em Fermi}). Since the start of operation in August 2008 the {\em Fermi} Large Area Telescope (LAT) has detected more than hundred thousand photons from  this object. Its brightness has been changing by at least two orders of magnitude (see Fig. \ref{fig:lc}): 
after a bright state in 2008, the blazar turned to a very low state in the spring of 2009, then the flux has been gradually increasing. 
The bright flares then occurred in November 2009 and  April 2010 and the brightest $\gamma$-ray flare ever observed  from a blazar happened in  November 2010.

The early data from the first 160 days of the {\em Fermi} observations have demonstrated a clear spectral break around 2.5--3 GeV \citep{Abdo09_3C454.3}. 
Similar breaks in the GeV range have been observed in several other 
FSRQs and low-synchrotron peak BL Lac objects \citep{Abdo10_blazars}. 
Later works by \citet{Fermi10_3C454} and \citet{Fermi11_3C454} dedicated to \source\ 
found the spectral hardness--flux correlation at a nearly constant break energy. 
First interpretations of the breaks invoking the external Compton mechanism with a truncated electron spectrum  \citep{Abdo09_3C454.3,FD10} do not reproduce as sharp a break as observed.  
The stability of the break energy favours instead the absorption of the GeV photons by the photon-photon ($\gamma\gamma$) 
pair production on the \mbox{He\,{\sc ii}} and \mbox{H\,{\sc i}} Lyman recombination continua (LyC) from  the broad-line region (BLR),
as proposed by \citet[][hereafter PS10]{PS10}.  
A relatively high opacity in the \mbox{He\,{\sc ii}} LyC observed in a number of bright blazars 
implies the location of the $\gamma$-ray emitting region within the highly ionized inner part of the BLR. 

Motivated by the extremely rich photon statistics from \source, in the  present work we particularly try to 
answer the following questions: 
\begin{enumerate}
\item  Is the spectral break consistent with the $\gamma\gamma$-absorption on the He\,{\sc ii} LyC at various source fluxes? 
We do not have a freedom of choosing the position of the break, with the only  free parameter being the LyC column density.
A good fit at a rich statistics would be a strong argument in favour of the absorption hypothesis.

\item Is there any correlation between the break sharpness  (i.e. absorption opacity in our interpretation) and the flux? 
If yes, then this would imply a varying location  of the $\gamma$-ray emitting region.

\item What is a proper characterization of the spectrum below the break?  In PS10 the hypothesis for the underlying (unabsorbed) spectrum was a power-law. The underlying spectrum can actually be slightly curved with respect to the power-law.
\end{enumerate}

\section{Data and their analysis} 
\label{sec:data}

The light curve of \source\ from the start of the {\em Fermi} data records to  2011 April 1 is shown in Fig. \ref{fig:lc}. 
We  split the data into ten time intervals covering periods of more or less constant flux levels  
and analyze the spectrum in each interval. 
We use our own software for the data analysis because the standard software distributed by the {\em Fermi} 
LAT team does not contain the required spectral models. 
The energy-dependent exposure function was calculated using the spacecraft pointing history: 
\begin{equation}
\mbox{Exposure}(E,\Delta T) = \int_{\Delta T} S(E,\theta(t)) \ \rmd t,
\end{equation}  
where $\Delta T$ is the time interval of interest, 
$\theta$ is the angle between the detector axis and the direction to the object and 
$S (E,\theta)$ is the detector effective area at energy $E$. 
The spectral models folded with the exposure function are fitted to the observed counts in 13 logarithmically spaced 
energy bins in the energy range 0.15--60 GeV. 

\begin{figure*}
\epsfig{file=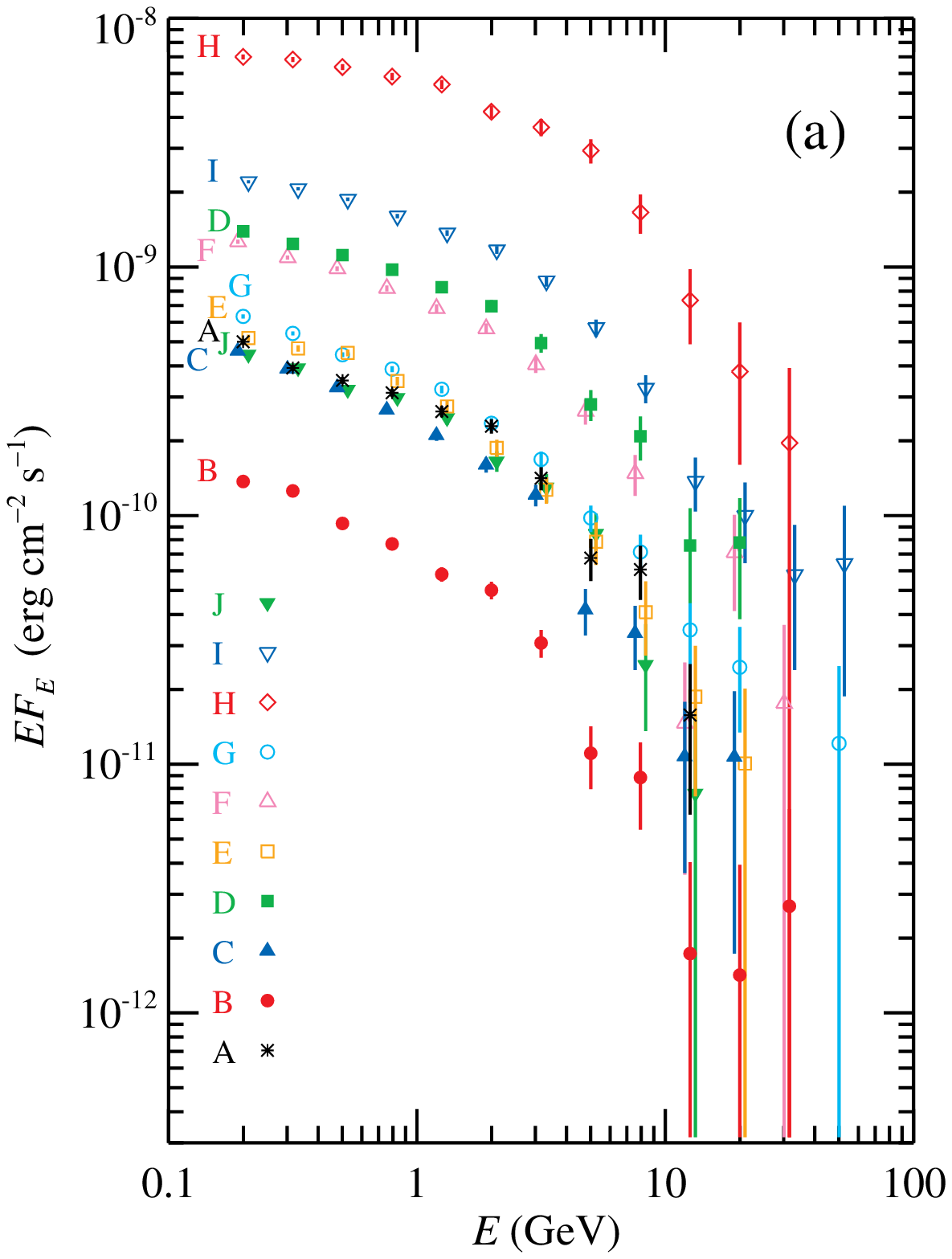, width=7.5cm}\hspace{1cm}
\epsfig{file=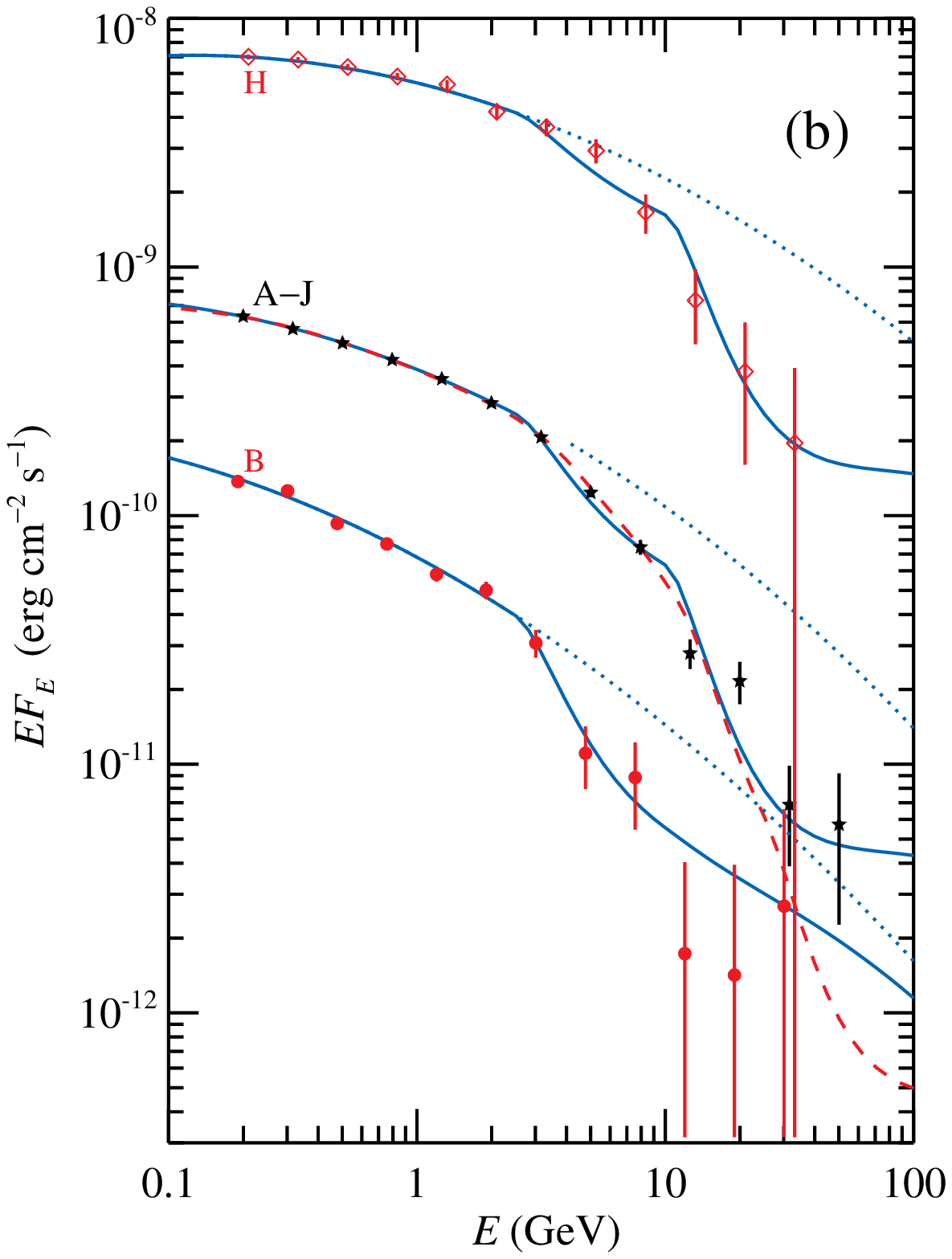, width=7.5cm}
\caption{(a) Spectral energy distribution (SED) of \source\  for ten intervals A--J.
(b) SED of \source\ for intervals B and H of the lowest and highest fluxes 
as well as averaged over the whole observation period of 2.5 years (sum of intervals A--J).
The solid curves show the best-fitting lognormal  ($\sigmaln=2.7$) + DA model. The dashed curve 
represents the fit with the lognormal distribution absorbed by BLR at a fixed ionization parameter $\log \xi=2.5$ (see PS10). 
The dotted curves show the unabsorbed lognormal distributions.  
} 	
\label{fig:all_spe}
\end{figure*}

We included the events of classes 3 and 4 and imposed the cuts at zenith angle $<$105$\degr$.  
We used \verb|P6_V3_DIFFUSE|  version of the response function.  
The background was measured in a circle with the radius 6$\degr$, separated from the object by $\sim$15$\degr$ and avoiding local sources. We accumulate counts in the circle centered at the source location with the energy-dependent radius $r=\max\{10\degr (E/100\ \mbox{MeV})^{-0.7}, 0\fdg 6\}$ (which corresponds to 95\% containment). 
Statistical errors were treated as Gaussian, except in a few bins at higher energies, where the number of photons is low, we use Poisson likelihood adding $-2\ln P(n,\mu)$ to $\chi^2$ (here $n$ is the number of counts in the bin and $\mu$ is the prediction of the model). The number of such bins is small and the  meaning of $\chi^2$ is not significantly affected. For the minimization we use the standard code {\sc minuit} from the CERN library. 
We checked that our software reproduces within statistical errors the results of the {\em gtlike} task 
for standard  spectral models such as power-law and broken power-law (compare table 2 in PS10 with table 1 in \citealt{Abdo10_blazars}).
Fig. \ref{fig:all_spe}(a) presents the spectra of \source\ accumulated in each time interval. 

\begin{table*}
%\begin{scriptsize}
\centering
\caption{Spectral properties of \source.  \label{tab:results} 
}
\begin{tabular}{@{}ccccccccccccc@{}}
\hline                            
{Interval} & Dates$^a$ &  \multicolumn{4}{c}{$\chi^2$/dof} & $\sigmaln$$^d$ & $F$$^e$  &  $E_{\rm peak}$$^f$   &   $\tau_{\rm He}$$^f$  &  $\tau_{\rm H}$$^f$  &  $\chi^2$/dof$^f$
\\
& MJD & Power-law    &  PL + DA$^b$ & Lognormal  & Logn + DA$^c$  &  &  & MeV   & &  & \\   
\hline
 A  		&  54684--54759 & 84/12    & 7.7/10 & 37/11   & 7.7/9    & $>$4$^g$ 	          & 2.8   & 32$\pm6$  & 3.2$_{-0.9}^{+1.0}$   & 21$_{-14}^{+29}$  & 16/10 \\
 B   		& 54759--55034 & 34/12    & 8.4/10 & 16/11    & 7.3/9   & 3.3$^{+0.9}_{-0.7}$  & 0.68  & 11$\pm3$  & 4.5$_{-1.2}^{+1.3}$  & 0.4$_{-0.4}^{+7.0}$ &    7.5/10           \\      
 C  		& 55034--55159 & 65/12    & 6.8/10 & 16/11   & 3.7/9    & 3.2 $\pm$0.6            & 2.4   & 15$\pm3$  & 4.0$\pm$0.9  & 3.1$_{-3.0}^{+8.3}$           & 3.8/10 \\   
 D  		& 55159--55199 & 81/12    & 6.7/10 & 14/11   & 2.9/9    & 3.1$^{+1.0}_{-0.4}$  & 9.1    & 49$\pm6$  & 2.8$\pm$0.7  & 4.6$_{-2.7}^{+3.0}$          & 3.2/10 \\
 E  	 	& 55199--55284 & 63/12    & 40/10  & 4.4/11  & 4.4/9    &  1.6$\pm$0.1            & 3.0   & 23$\pm4$  & 4.0$\pm$1.0 & 3.9$_{-3.9}^{+5.5}$           &  19/10 \\
 F  	 	& 55284--55344 & 112/12  & 23/10  & 27/11   & 17/9     & 2.9$^{+0.5}_{-0.3}$   & 7.6   & 33$\pm5$  & 2.6$\pm$0.6  & 4.0$_{-2.1}^{+2.7}$           &  17/10  \\
 G  		& 55344--55517& 85/12    & 13/10  & 18/11   & 7.9/9    & 3.2$^{+0.7}_{-0.5}$   & 3.4   & 19$\pm3$  & 2.1$\pm$0.6  & 3.1$_{-2.0}^{+2.4}$            &  8.8/10 \\
 H               	& 55517--55522 & 79/12    & 18/10 & 11/11   & 5.9/9     & 2.6$^{+0.2}_{-0.3}$  & 55     & 126$\pm17$  & 1.6$\pm0.6$          & 7.3$_{-2.5}^{+2.8}$  &   6.8/10         \\
 I            	& 55522--55587 & 242/12  & 33/10  & 28/11   & 10/9    & 2.5$\pm$0.2            & 15      & 60$\pm5$  & 2.7$\pm$0.4             & 5.2$_{-1.5}^{+2.6}$         &   11/10 \\
 J              	& 55587--55652 & 43/12  & 11/10  & 13/11   & 7.2/9    & 2.4$^{+0.6}_{-0.4}$    & 2.6    & 26$\pm6$  & 3.1$_{-1.1}^{+1.3}$  &  $>$8$^g$                &    7.3/10        \\
A+C+E+G 	& & 315/12  & 33/10 & 42/11   & 9.0/9     & 2.7$\pm$0.2            & 2.9     & 20$\pm2$  & 3.2$\pm0.4$  & 6.2$_{-1.8}^{+2.2}$          &  9.0/10 \\
D+F+I		& & 565/12  & 56/10 & 71/11   & 23/9      & 2.8$^{+0.3}_{-0.2}$ & 11    & 48$\pm3$  & 2.7$\pm0.3$         & 7.2$\pm$1.5  &  23/10  \\ 
A--J    & 54684--55652 & 1099/12 & 85/10 & 129/11& 23/9      & 2.6$\pm$0.2           & 3.9      & 35$\pm2$  & 2.7$\pm0.2$         & 8.1$_{-1.2}^{+1.5}$         &  23/10  \\
\hline
      \end{tabular}
\begin{flushleft}{
  $^{a}$The start and the end of the interval. 
  $^{b}$Power-law plus double absorber  (DA) model. 
  $^{c}$Lognormal plus  DA model. 
    $^{d}$Best-fitting values of $\sigmaln$ for the {lognormal + DA} model.   
  $^{e}$Flux $EF_E$ at 1 GeV in units of 10$^{-10}$ erg cm$^{-2}$ s$^{-1}$.  
   $^{f}$Best-fitting values of $E_{\rm peak}$,  $\tau_{\rm He}$ and  $\tau_{\rm H}$ and the corresponding $\chi^2$/dof
    for the {lognormal (with $\sigmaln=2.7$) +  DA} model.   $^{g}$1$\sigma$ lower limit.  
}\end{flushleft} 
%\end{scriptsize}
\end{table*}

\section{Spectral fits} 
\label{sec:fits}

First, we fit the data with a simple power-law model. This is our first null-hypothesis against which we check the significance of the break. 
The resulting values of $\chi^2$ are given in  column 3 of Table \ref{tab:results}. 
Then we add the $\gamma\gamma$-absorption by photons in two ``lines'' at  54.4 and 13.6 eV 
(double absorber model, DA), which correspond to the energies of the \mbox{He\,{\sc ii}} and H 
LyC, which are the strongest features in the BLR spectra at relevant energies.
These  continua are basically line-like because  the photoionized region temperature is typically 
much below the corresponding ionization potentials of H and He.
Within this model, the spectral breaks at expected at the energies $261/(E_{\rm line} ({\rm eV}) [1+z])$ GeV 
corresponding to the threshold of the $\gamma\gamma$ absorption on  photons of energy $E_{\rm line}$ and redshifted. 
In \source, the breaks should appear at 2.6 and  10.3 GeV, which is consistent with observations 
(\citealt{Abdo09_3C454.3, Fermi10_3C454}; PS10; see Fig.~\ref{fig:all_spe}).

The absorption strength  by the isotropic line photons is parameterized by the Thomson optical depth (see PS10): 
\begin{equation} \label{eq:taut}
\tau_{\rm T} =  N_{\rm ph}  \sigma_{\rm T}  
\approx 110 \ \frac{L_{\rm line,45} }{R_{18}} \frac{10\ {\rm eV}}{E_{\rm line}} ,
 \end{equation} 
where $N_{\rm ph}$ is the photon column density along the line of sight,  $\sigma_{\rm T}$ is the Thomson cross-section, 
$L_{\rm line}$ is the line luminosity, and $R$ is the typical size. (We defined $Q=10^xQ_x$ in cgs units.)
The energy-dependent opacity $\tau_{\gamma\gamma}(E,E_{\rm line})$ 
is obtained by multiplying $\tau_{\rm T}$ by the angle-averaged 
$\gamma\gamma$ cross-section in units of $\sigma_{\rm T}$,  
which has a maximum of $\sim$0.2 at three times the threshold energy (see \citealt{Aha04}; PS10).   
The opacities produced by two ``lines'', $\tau_{\rm He}$ and $\tau_{\rm H}$, were free parameters. 
We see that with the power-law null-hypothesis all spectra have very significant breaks and require non-zero absorption opacities (column 4 in Table \ref{tab:results}). 
However, the residual value of $\chi^2$ is still high for many spectra, especially 
for the average spectrum corresponding to the entire observational period, 
mostly because of the spectral curvature below the break. 

The simplest way to account for this curvature is to replace the power-law null-hypothesis by the lognormal distribution (or ``log-parabola'', which is essentially the same function in a different representation, see 
\citealt{Massaro04},   \citealt{Fermi10_3C454}), which we use  in the decimal form 
\begin{equation}
EF_E \propto 10^{-\log^2 (E/ E_{\rm peak}) /\sigma_{\rm ln}^2} . 
\end{equation}
The resulting fits are shown in columns 5 (without absorption) and 6 (with DA) of Table \ref{tab:results}. 
We see a significant decrease in $\chi^2$ compared to the power-law-based models. 
The DA model gives a substantial reduction  in $\chi^2$ in all cases  (except for interval E). 
These fits are superior compared to any simple phenomenological model considered earlier (e.g. lognormal, 
broken power-law, or an exponentially cutoff power-law, see \citealt{Fermi11_3C454}).

If the spectral break is indeed caused by the $\gamma\gamma$-absorption on the \mbox{He\,{\sc ii}} LyC,
its position should be stable. 
In that case we can expect that the break should become more significant 
in the average spectrum corresponding to the whole observation period. 
Indeed, this is the case, the lognormal distribution without absorption fits the data  (see Table \ref{tab:results}) 
at $\chi^2 = 129$ for 11 degrees of freedom (dof). 
The DA improves the fit (see Fig. \ref{fig:all_spe}(b)), 
but the residual value $\chi^2/\mbox{dof} = 22.8/9$ is still high. 
The absorption model for the BLR at a fixed ionization parameter $\log \xi=2.5$ (see PS10)
gives $\chi^2/\mbox{dof}=27/10$,  with the quality of the fit being better ($\chi^2/\mbox{dof}=7/7$) in the range $<$15 GeV. 
The main contribution to the residual $\chi^2$ comes actually from the high-energy tail of the spectrum and might have a simple interpretation: the absorber optical depth varies during the observation and the sum of the spectra with low and high opacities differs from the spectrum at a fixed intermediate opacity.

\begin{figure}
\epsfig{file=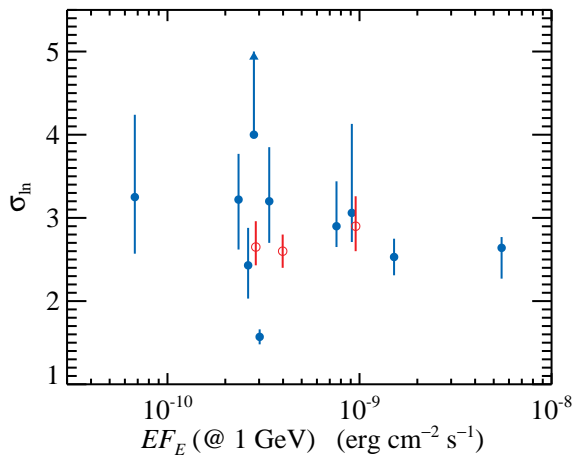, width=8cm}
\caption{Dependence of the width of the best-fitting lognormal + DA model on flux at 1 GeV. 
The parameters for the intervals A+C+E+G, A--J and D+F+I  are shown by open circles. } 	
\label{fig:sigma}
\end{figure}

\begin{figure}
\epsfig{file=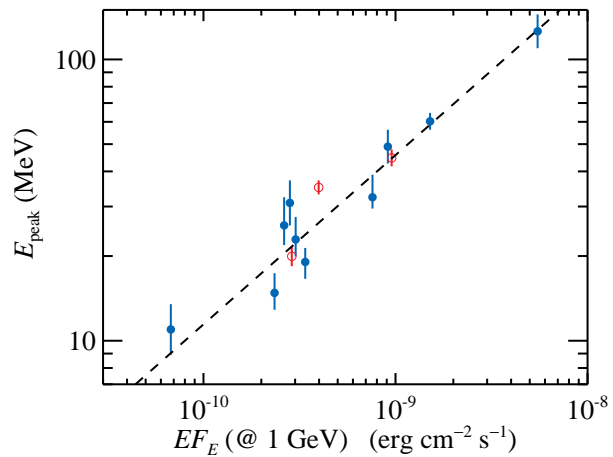, width=8cm}
\caption{
Dependence of the peak (in $EF_E$) of the best-fitting lognormal  
(with $\sigmaln=2.7$) + DA model on flux at 1 GeV. 
The dashed line shows the best-fitting relation for intervals A--J 
given by Equation (\ref{eq:epeak_fit}).
The parameters for the intervals A+C+E+G, A--J and D+F+I  are shown by open circles. 
} 	
\label{fig:epeak}
\end{figure}

\begin{figure}
\epsfig{file=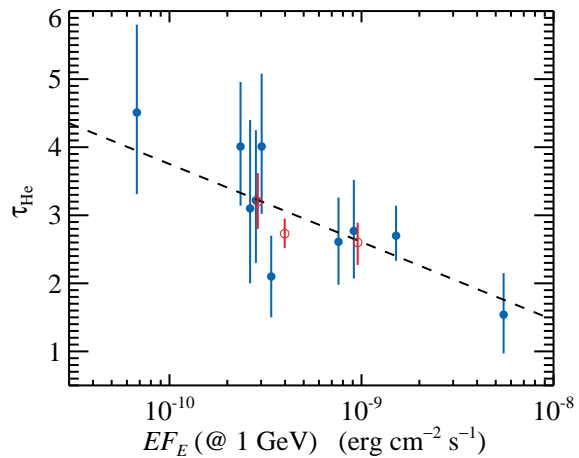, width=8cm}
\caption{Dependence of the Thomson optical depth 
in the \mbox{He\,{\sc ii}} LyC (defined by Equation (\ref{eq:taut})) on flux at 1 GeV. 
The model and notations are the same as in Fig.  \ref{fig:epeak}.
The dashed line shows the best-fitting relation for intervals A--J given by Equation (\ref{eq:tauHe_fit}).
 }
\label{fig:tauHe}
\end{figure}

The best-fitting value of $\sigmaln$ is nearly constant $\sim 2.5$--3 
(see column 7 of Table  \ref{tab:results} and Fig. \ref{fig:sigma}), with the interval E being again an outlier.
In order to avoid extra statistical noise, we fix $\sigmaln = 2.7$ and then probe the correlation 
of other parameters with the source brightness (see Table~\ref{tab:results}).  
We find a highly significant  correlation of the peak energy $E_{\rm peak}$ with flux, 
which is well represented by a power-law (Fig. \ref{fig:epeak}): 
\begin{equation} \label{eq:epeak_fit}
E_{\rm peak} = (46\pm2) \ F_{-9}^{0.60\pm0.04}\ \mbox{MeV},
\end{equation}
where $F=EF_E$ at 1 GeV. This  confirms a previously known hardness--flux correlation  \citep{Fermi10_3C454}. 
We also observe a marginally significant (2.5$\sigma$)
 anti-correlation between the $\gamma\gamma$-opacity by the \mbox{He\,{\sc ii}} LyC and the flux  (see Fig. \ref{fig:tauHe}):
\begin{equation} \label{eq:tauHe_fit}
 \tau_{\rm He} = (2.61\pm0.22) - (1.15\pm 0.47)\ \log  F_{-9}.
\end{equation}
The opacity $\tau_{\rm H}$ in the \mbox{H\,{\sc i}} LyC does not vary significantly.

\section{Discussion and summary}

The  BLRs around quasars emit a number of  lines associated with very different ionization stages, 
with the high-ionization lines being broader \citep{PW99}.   
Reverberation mapping demonstrated a strong anti-correlation between the line width and the time delays of the line 
response to the continuum variations \citep{PW00}. 
All this implies strong stratification of the BLR that extends over two orders of magnitude in distance \citep{Krolik99,AGN2}.
The BLR size, as measured from the \mbox{C\,{\sc iv}}~$\lambda$1549 line delays, 
scales with the accretion luminosity as 
$R_{\ \mbox{\scriptsize C\,{\sc iv}}} \approx 0.2 L_{47}^{1/2}$ pc \citep{Kaspi07}. 
However, the delays in the high-ionization He\,{\sc ii}~$\lambda$1640 line are 3--5 times smaller 
in Seyferts \citep{Korista95,PW99},  while the Balmer lines give sizes 2--3 times larger. 
As the opacity in the GeV range depends on the line compactness  ($L/R$), 
models assuming that all lines are produced 
at the same distance from the central source \citep[e.g.][]{Liu06,Reimer07},  strongly underestimate the 
$\gamma\gamma$-opacity by the relatively weak, high-ionization lines produced in a small volume
(if the $\gamma$-rays are produced closer to the centre)
and overestimate the opacity for multi-GeV photons from the strong, low-ionization lines which 
are produced much further away than the fiducial distance obtained from  \mbox{C\,{\sc iv}}.

A high average $\gamma$-ray luminosity of \source\ during the last two years 
with flares up to $L_{\gamma}\sim 2\times 10^{50}$~erg~s$^{-1}$  \citep{Fermi11_3C454} 
provided an order of magnitude increase in photon statistics compared to the earlier data analysed by 
\citet{Abdo09_3C454.3} and PS10.  In became even more evident that the spectral breaks at $\sim$3 GeV are real.
The spectra at all fluxes are well described 
by a lognormal distribution absorbed  
by the \mbox{He\,{\sc ii}}  and \mbox{H\,{\sc i}} LyC at 54.4 and 13.6 eV.
The fits with this model are superior compared to the previously used simple phenomenological models,  
especially at high fluxes with high photon statistics. 

The BLR size corresponding  to the He\,{\sc ii} and H\,{\sc i}  LyC
can be estimated from Equation (\ref{eq:taut}): 
\begin{equation} 
R_{\rm He} \approx 0.22\ L_{\rm He,44}  (\tau_{\rm He}/3) \ \mbox{pc} , \quad 
R_{\rm H} \approx 3.7\ L_{\rm H,45}  (\tau_{\rm H}/7) \ \mbox{pc} .
\end{equation}
\citet{Wills95} report the luminosity of \source\ in Ly$\alpha$ of $\sim10^{45}$~erg~s$^{-1}$ and 
in He\,{\sc ii}~$\lambda$1640
%(i.e. H$\alpha$) ~
of $6\times10^{43}$~erg~s$^{-1}$. 
Assuming equal luminosities in LyC and  Ly$\alpha$, the resulting 
$R_{\rm H}$ is much larger than the usually quoted sub-pc size of BLR, but is consistent 
with the absence of Ly$\alpha$ variability in powerful quasars \citep{Kaspi07}. 
The He\,{\sc ii} LyC luminosity is probably larger than $10^{44}$~erg~s$^{-1}$,
but is model dependent and therefore the estimate of $R_{\rm He}$ is less certain. 
A rather large ratio $\tau_{\rm He}/\tau_{\rm H}$ (between about 1/4 and 1) 
and the observed $\tau_{\rm He}$--flux anti-correlation 
indicate that the $\gamma$-ray emission region lies close to the boundary  of the He\,{\sc iii}  
zone at $R\lesssim R_{\rm He}$ and moves out at higher luminosity. 
Absorption troughs at 0.3--0.7 GeV due to the high-ionization C and O lines 
are potentially detectable, but require high photon statistics and better calibration of LAT. 

We note that the typical values of $\tau_{\rm He}\sim 2$--4 imply that the maximum optical depth 
due to the He\,{\sc ii} LyC photons 
(at about 7 GeV -- three times the threshold value of 4.8/$(1+z)$ GeV) varies around 0.4--0.8. 
The absorption by \mbox{He\,{\sc ii}} LyC is very significant and accurately describes the data in the 3--15 GeV range. 
At high energies, the opacity decreases roughly inversely with the photon energy and becomes negligible 
above $\sim$300 GeV. The opacity there is dominated by the optical/IR lines which are mostly 
produced in the low-ionization zone of BLR at large distances. This strongly reduces the opacity and can
make the BLR transparent for the multi-GeV to TeV photons.  

A probable excess of  photons above 15 GeV (especially for the average spectrum) 
over the prediction of the model with the best-fitting absorption indicates that 
the spectrum is a superposition of emission states with different opacities 
due to the extended or moving emission zone. 
This proposal is supported by the spectral variability seen 
during the latest flare in November 2010 \citep{Fermi11_3C454}, 
where photons above 10 GeV mostly arrive at the end of the flare, clearly 
indicating the position of the $\gamma$-ray emitting region further away from the black hole.  

Our findings that the spectral peak of the underlying continuum strongly correlates with the flux, with 
the  width of the lognormal distribution  being roughly constant, indicate a more efficient acceleration at high fluxes. 
These variations might not be consistent with the interpretation of the spectral curvature as the Klein-Nishina reduction 
of the Compton scattering efficiency of the Ly$\alpha$ photons  \citep{Fermi10_3C454}. 
This in turn would imply that  the seed photon energy density is not dominated by 
the hydrogen Lyman continuum and line photons 
(in the high-ionization zone of the BLR  the metal lines and the \mbox{He\,{\sc ii}} Lyman photons  are actually  
more energetically important).
Alternatively, the steady electron distribution is modified by inefficient cooling  in the Klein-Nishina regime \citep{ZK93,MSC05}. 

The GeV breaks observed in bright blazars 
still do not answer the question what is the $\gamma$-ray production mechanism. 
It can be Compton scattering of the accretion disc  \citep{DS93} or BLR photons \citep{SBR94}, or 
even the synchrotron radiation of relativistic pairs. The efficient means for particle acceleration 
can be provided by the photon breeding mechanism \citep{StP06,SP08}.
 
A small distance  to the beginning of the $\gamma$-ray emitting region 
is supported by the fast variability observed from \source\ \citep{TGB10,Bonnoli11}.
Association of the beamed $\gamma$-ray emission with a relativistic jet implies that the jet is already accelerated at a sub-parsec distance from the black hole 
(i.e. at thousand Schwarzschild radii from the $\sim$10$^9 M_{\odot}$ black hole, \citealt{Bonnoli11}).
This, in turn, argues against the jet launching mechanism by the accretion disc \citep{BP82}, with  
the only known alternative being the extraction of the black hole spin energy by the Blandford-Znajek process 
\citep{BZ77,2007MNRAS.380...51K}.
 
\section*{Acknowledgments}
This research was supported  by the Academy of Finland grant 127512  and the Magnus Ehrnrooth foundation.
The research made use of public data obtained from the Fermi Science Support Center. 
We thank Gabriele Ghisellini, Chuck Dermer and Maxim  Barkov for useful discussions. 

%\bibliographystyle{mn2e}
%\bibliography{../allbib}
%\end{document}

\label{lastpage}

\end{document}